# Spot Focusing Coma Correction by Linearly Polarized Dual-Transmitarray Antenna in the Terahertz Region

Ka Kit Kelvin Ho, *Student, IEEE,* Geng-Bo Wu, *Member, IEEE*, Bao-Jie Chen, *Member, IEEE,* Ka Fai Chan*, Member, IEEE, and* Chi Hou Chan, *Fellow, IEEE*

*Abstract*—**Focus scanning is critically important in many terahertz (THz) imaging and sensing applications. A traditional single focusing transmitarray can achieve a good focus when the source is on-axis but moving the source off-axis produces a significant aberration. This paper presents a novel approach to reducing coma in off-axis scanning in the THz region. Here, a dual transmitarray solution is proposed, in which a transmitarray with an optimized phase profile is placed behind a regular phase profile transmitarray. A linearly polarized, dual-transmitarray antenna was fabricated for validation, and the focusing performances were experimentally characterized. The measured results are in good agreement with the theoretical ones. The generated spot of the dual-transmitarray antenna remains focused on an angle up to 50°, with a -3 dB spot size of less than 4 mm at 290 GHz. The measured near-field sidelobes are all below -10 dB within the whole scanning range.**

## I. INTRODUCTION

Terahertz (THz) waves are increasingly popular for their numerous applications, ranging from wireless communications [1] to medical scanning [2], [3]. It has a shorter wavelength than microwave, allowing scanning applications involving THz waves to have a far greater resolution while being far less harmful to biological substances than X-ray [4]. One of the most exciting developments is using THz waves for scanning applications, such as in standoff personnel detection at security checkpoints [5], [6]. The unique absorption frequencies of substances also allow THz waves to be used in medical scanning applications [4]. THz waves would have to be focused on tiny sections of a sample in both cases. As such, a near-field focus point can be raster-scanned to gather information about the structure of a medical sample. As a result, it is paramount to have a tight and scanning focus spot that can be used to 'pinpoint' features in the sample [7].

Scanning with the use of THz waves necessitates near-field focus steering. While electrical scan techniques have been devised at much lower microwave frequencies, traditional mechanical scan techniques remain popular in the THz band due to their simplicity, low cost, and ease of fabrication [8]. While many spot-focusing antennas have been designed to achieve a focusing effect, traditional designs often lead to defocusing when achieving off-axis focus [9]. Designs involving reflectors or dielectric lenses often employ a parabolic phase profile. While this succeeds a tight focus in producing a focus spot on-axis, the off-axis focus often introduces coma into the resulting focus spot. Typically, when the off-axis scanning angle exceeds 30º, artifacts such as the broadening of intensity peaks, pattern degradation, and comatic aberrations would be presented. A broadened focus spot would reduce resolution and accuracy as the resolution depends on the diffraction-limited focus spot [10]. The presence of high-intensity sidelobes or pattern degradation would be undesirable in most scanning applications.

Comatic aberrations, or coma, are common forms of aberration arising from a lens's imperfection [10], [11]. As off-axis rays have varying focal lengths, they cannot converge to a tight focus. Consequently, point sources that are located off-axis appear distorted. Antenna design involving parabolic phase profiles includes transmitarray [12] and reflectarrays [13]. Transmitarrays are antennas with a feed source illuminating a thin, transmitting surface, which can be constructed with metasurfaces or a similarly phase-shifting surface [14], [15]. The feed source would be located at the equivalent of the focal point [12]. Similarly, a reflectarray operates on the same principle but instead reflects the waves from a feed source. Some near-field focusing transmitarrays [16] and reflectarrays [17] have been reported in the open literature. As transmitarray and reflectarray antennas with a parabolic phase profile are functionally identical to spherical lenses in classical optics, they share the common problem of introducing comatic aberration when performing off-axis focusing.

Traditional approaches generally only utilize one reflectarray or transmitarray to perform focusing, and suffer from off-axis aberration [16]–[18]. In classical lens optics, a solution to

This work was supported by the Hong Kong Research Grants Council under the GRF grant (Project No. CityU 11217720).

K. K. K. Ho was affiliated with the State Key Laboratory of Terahertz and Millimeter Waves, City University of Hong Kong (CityU), Hong Kong, China, at the time of this project. He is currently affiliated with University College London, Department of Physics and Astronomy, London, United Kingdom.

G. B. Wu, B. J. Chen, and K. F. Chan are with the State Key Laboratory of Terahertz and Millimeter Waves, City University of Hong Kong (CityU), Hong Kong, China (email: bogwu2@cityu.edu.hk).

C. H. Chan is with the State Key Laboratory of Terahertz and Millimeter Waves, Department of Electrical Engineering, CityU, Hong Kong, China (e-mail: eechic@cityu.edu.hk).

rectify such a problem utilizes methods such as introducing an aperture stop or passing the beam through several lenses, each correcting for different aberrations [19]. Inspired by such an approach, this paper introduces a 'doublet' setup to reduce comatic aberration in off-axis scanning. The first transmitarray would have a phase profile optimized to minimize the coma via altering the off-axis rays' focal length. The beam from the first transmitarray is then passed through an additional focusing transmitarray which refocuses the beam, allowing for a focus with a significantly reduced coma at the imaging plane.

This paper considers a system fed with a linearly polarized source in which only the source would be mechanically steered. Transmitarrays would be used to focus the waves from the source. This is a setup that has been established [20], and is also used in some of this group's work [8], [21] but for one-transmitarray and far-field applications. While it is possible to steer the transmitarray or whole antenna systems instead [13], [22], source steering is preferred because applications, such as standoff personnel screening, use bulky reflectors/transmitarrays/lenses focus a source, in which mechanically steering the reflectors/transmitarrays/lenses or whole antenna system are quite difficult and cumbersome. Furthermore, such applications require a rapid scan to be performed in one second [6], [23]. This requires the focusing aperture to be mechanically steered at high speeds, which can be quite difficult for heavy and cumbersome reflectors/transmitarrays/lenses. Furthermore, medical applications such as *in vivo* disease diagnosis demand handheld and fast-scan portable THz imaging systems [3]. This leads to considering a movable source, as THz sources are increasingly miniaturized. There have been many promising developments in making small THz sources through the use of IC technology [24], [25]. This system considers a 1D scanning system that can be used to construct portable THz scanning apparatus that can be easily carried to various venues to provide medical scanning or safety scans [5].

## II. THEORETICAL MODEL

This section describes a theoretical model constructed to calculate the resultant field radiated by the transmitaray at the imaging plane given a source distribution. The transmitarray would simply be treated as a component which can alter the phase of the incoming electromagnetic (EM) wave, and its specific construction would not be taken into account. This will be subsequently used in Section III which optimizes transmitarray designs.

In this work, the Angular Spectrum Method (ASM) [10] would be used to simulate, optimize and design the transmitarrays. A temporal dependence of $e^{-j\omega t}$ is assumed and suppressed in notation for convenience throughout the paper. The field in the imaging plane $E$ and source plane $E_0$ are related by:

$$FT[\mathrm{E}] = FT[\mathrm{E}_0]e^{j\left(\sqrt{k_0^2-k_x^2-k_y^2}\right)z} \tag{1}$$

where $k_0$ is the free-space wavenumber at the operating frequency of 290 GHz, and $k_x$ and $k_y$ are the transversal wavenumbers along the $x$- and $y$-axis, respectively. The Fourier transform of initial field distribution in the source plane is represented by $FT[\mathrm{E}_0]$, and $e^{j\left(\sqrt{k_0^2-k_x^2-k_y^2}\right)z}$ represents the spatial transfer function. The transmitarray is expressed as a transparency function, which is a combination of an amplitude function and phase function, which modifies the EM wave's amplitude and phase distribution:

$$t(x,y) = A(x,y)e^{j\phi(x,y)} \tag{2}$$

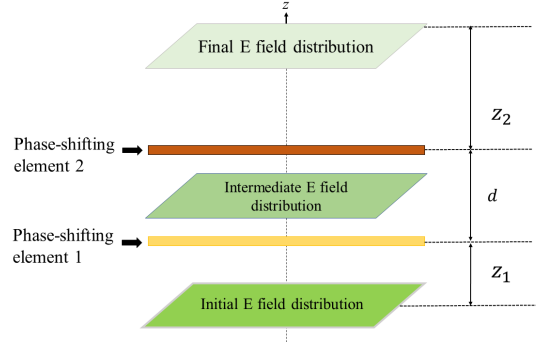

Fig. 1. Flowchart shows how the theoretical model performs wave propagation.

The EM field distribution originating from the source would be propagated through free space in the form of (1) and then the transparency functions in the form of (2). This is the approach taken by the theoretical model to calculate how the wave propagates [26].

On any setup, the theoretical model proceeds as follows. First, the initial field distribution would be defined, as shown in Fig. 1. This could be the amplitude and the phase information of a horn source, taken at the correct operating resolution and size. Then this would be propagated through $z = z_1$ to reach the transmitarray, which is performed via taking the Fourier transform of the field at $z = 0$, multiplying it with the spatial transfer function, and then taking the inverse Fourier transform of it using Eq. (1) [27]. Subsequently, the propagation is simulated by multiplying the field distribution with the transparency function $t(x,y)$ in (2), before propagating through $z = z_2$ to the imaging or focal plane. Such a setup can easily be extended to calculate the field after propagating it through multiple transmitarrays or across different distances.

The ASM adopted in this manuscript has been used for calculating the near-field distributions for near-field planes [28], [29] and lens [10], [26]. Therefore, the ASM is adopted to calculate the near-field distribution of transmitarray even if the dual-transmitarrays are in the near-field zone of each other, considering that the distance between the two transmitarrays is smaller than $2D^2/\lambda$, where $D$ is the diameter of the transmitarray.

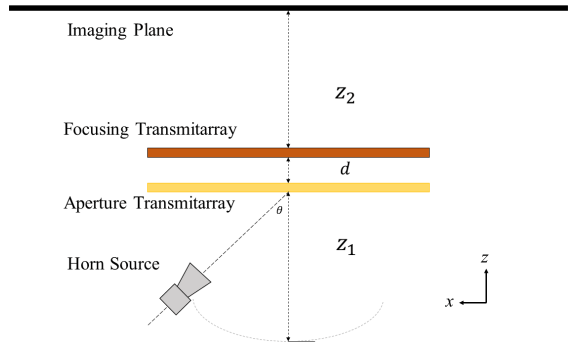

Fig. 2. Schematic of transmitray doublet to perform wide-angle focus scanning.

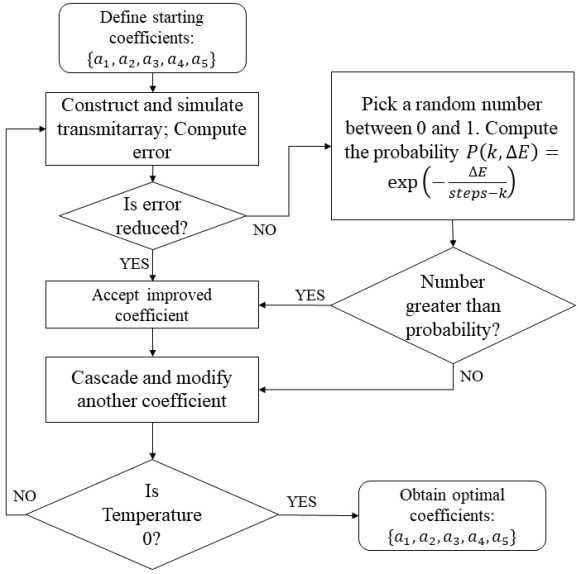

Fig. 3. Flowchart shows the optimization process based on a modified simulated annealing algorithm.

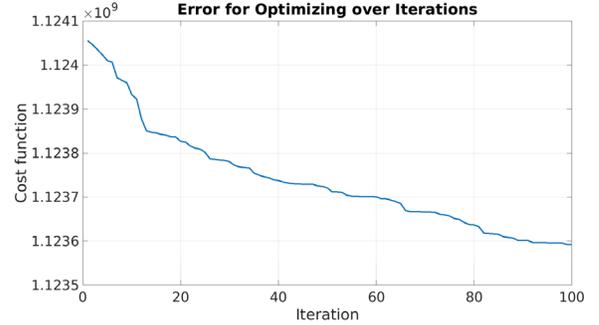

(a)

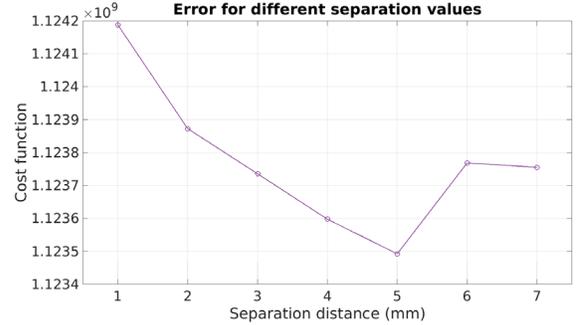

(b)

Fig. 4. Optimization results. (a) The cumulative cost function across all angles as a function of iteration. (b) Cumulative cost function after optimization with different values of interplane separation $d$.

## III. TRANSMITARRAY DOUBLET DESIGN

### A. Double Transmitarry

The configuration of the proposed doublet transmitarray is shown in Fig. 2, which consists of an aperture transmitarray and a focusing transmitarray, illuminated by a feed source moving in a circular arc. In this case, the focusing transmitarray has a phase profile given by [30]:

$$\phi_f = k_0(R_1 + R_2) \qquad (3)$$

where $R_1$ and $R_2$ are the spatial distances from the feed source to the focusing trasnmitray element, and the spatial distance from the element to the focal point for the on-axis focusing scenario, respectively.

The aperture transmitarray has a phase profile which can be expressed as a polynomial up to a particular order [19]:

$$\phi_a = \sum_{n=1}^{N} a_n \left(\frac{r}{\rho}\right)^{2n} \qquad (4)$$

Here, $a_n$ are coefficients to be determined through an optimization algorithm. The variable $r$ represents the radial distance from the center of the aperture transmitarray, and $\rho$ represents the total radial length of the aperture transmitarray. An optimized transmitarray of the following phase profile allows for the correction of coma via adjusting the 'focal length' of the wave that passes through each part of the transmitarray. Such an adjustment results in a more uniform focal length on the area of focus, achieving a diffraction-limited focus. As mentioned above, the source moves in a circular arc located at a constant radius of $z_1$ behind the aperture transmitarray. The location of the source can be identified with the angle subtended by the arc from the axis $\theta$, as shown in Fig. 2. While a setup with regular transmitarray produces significant coma, the doublet setup can reduce the coma dramatically. This setup has been tested with $z_1$ being 56 mm and $z_2$ being 40 mm. Such a setup can correct for coma up to an angle of $50°$.

In general, the transmitarray design is principally similar to lenses. Like lenses, it is defined to be a radially symmetric thickness profile. As such, any thickness profile can be described by a polynomial function of $r$, in which $r$ is the radial distance from the center of the lens as given in (4). The transmitarrays constructed here will be circular and has a radius of $\rho = 10$ mm, corresponding to $9.3\lambda$ at 290 GHz. While the polynomial can be in any order, it must only contain even-number order terms to maintain a radially symmetric profile. Furthermore, an arbitrary cut-off of 5 terms, or up to the order of 10, is considered, as this allows for more efficient optimization.

## B. Optimization

An optimization procedure is used to adjust the individual coefficients of (4) to obtain an aperture transmitarray which can best reduce the effects of comatic aberration. The array of coefficients would subsequently be referred to in the paper as aperture coefficients.

The procedure for optimization is as follows. First, an initial set of aperture coefficients is defined. An initial or benchmarking attempt would utilize an array of zeros as a starting set of coefficients. These coefficients would be used to construct an aperture transmitarray, and the wave propagation would be performed from the source, through the constructed aperture transmitarray and the focusing transmitarray to the imaging plane using the ASM method introduced in Section II.

The cost function is defined as the discrepancy between the simulated image and an ideal distribution $M(x, y)$. The ideal distribution is a perfectly on-axis focused spot shifted to the point on the imaging plane in which the off-axis beam should be on. The ideal distribution's near-field sidelobe level (SLL) is set as $L$. When the realized SLL is smaller than this value, it does not contribute to the cost function. As the performance for all angles has to be accounted for, the cost function adds the error resulting from each scanning angle. Thus, the cost function can be written as:

$$
Cost = \sum_{angles} [\sum_{(x,y) \in focus\ zone} \left(F(x,y) - M(x,y)\right)^2
$$
$$
+ P \sum_{(x,y) \notin focus\ zone\ and\ |F(u,v)| > L} (F(x,y) - L)^2]
$$

(5)

The next step in the optimization process then considers altering the aperture coefficient. The constant is adjusted by adding or subtracting a random number between 0 and 1. Simulated annealing is adopted here to speed up the optimization. The 'learning rate' can also be adjusted by multiplying this number with a constant, which adjusts how fast the constant changes, thus affecting the rate at which the coefficients converge to an optimal solution. A new aperture transmitarray would then be created, and the wave propagation would be calculated using the developed ASM method, and the cost function would again be calculated. This fitness function would be compared with the one obtained before the coefficients' adjustments. If the cost function reduces, the adjustment will be accepted. If the cost function increases, the adjustment is accepted by an algorithm based on simulated annealing [31]. At every step, a temperature is calculated by $T = steps - k$, with $k$ being the current step. The probability of accepting an adjustment is $P(T, \Delta E) = \exp\left(-\frac{\Delta E}{T}\right)$, where $\Delta E$ is the error. Thus, there is a non-zero probability of accepting an adjustment even if it increases the error. However, the likelihood of accepting such an adjustment reduces as the iteration increases or the temperature decreases.

When all the aperture coefficients have been adjusted, the optimization process iterates onto the next step, in which the temperature reduces, and each of the aperture coefficients are adjusted again. This is iterated over the predefined number of steps, and the aperture transmitarray is optimized for that specific angle. The source field shifts to one that represents moving the incident beam to the next incident angle. The temperature is reset to the starting temperature, and the optimization continues to fit the aperture coefficients to minimize the aberrations caused by the new source. Note that the aperture coefficients do not reset, and the results from optimizing the source at another angle are used directly. A flowchart demonstrating the algorithm is summarized in Fig. 3.

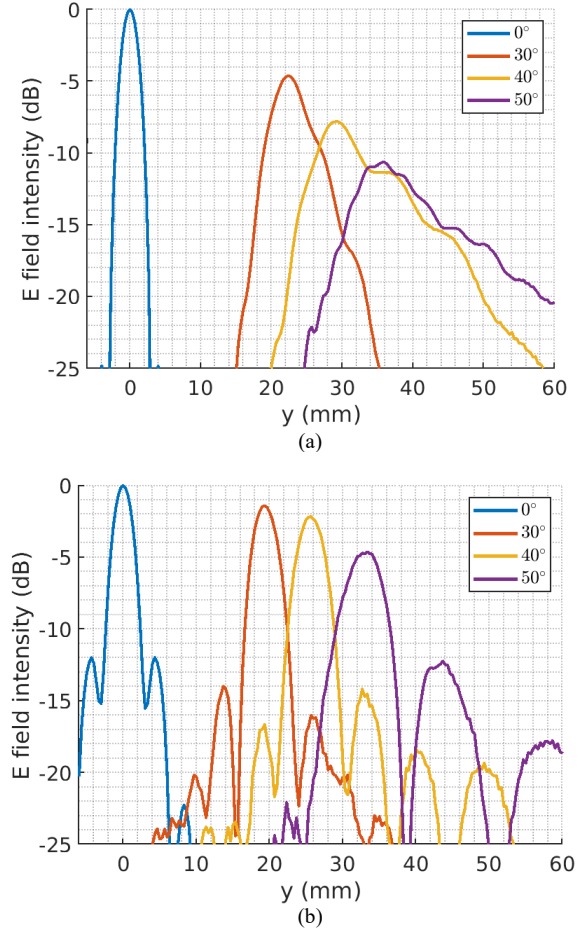

Fig. 5. One dimensional cross section of the intensity distribution generated by the theoretical model across the yz with different feed source rotation angles, for (a) the conventional parabolic phase-shifting transmitarray, and (b) the proposed doublet transmitarray.

In the actual implementation of the optimization for this setup, a simulation area, or the field area that is to be simulated, is 200 mm × 200 mm. The unit cell is one third of the wavelength which is 0.33 mm, and the plane has 601 × 601 unit cells, with the transmitarray spanning 2808 out of the total of 361201 unit cells. The source field distribution is given by the measured field from a horn antenna at a radius of 56 mm away from the aperture transmitarray and offset at 0°, 30°, 40°, and 50°. Angles smaller than 30° was omitted because a source displacement at those angles would only have minor comatic aberrations. Testing it at 0° ensures that the transmitarray's efficacy remains comparable to traditional transmitarrays. The number of steps iterated is determined to be 100 steps, as the simulation plateaus at a level of error after iterating for such number of steps. The cost function over the iteration steps is displayed in Fig. 4(a). On average, this requires a computational time of 550 seconds using a computer with 16-GB memory and Intel Core i7 CPU @1.80 GHz.

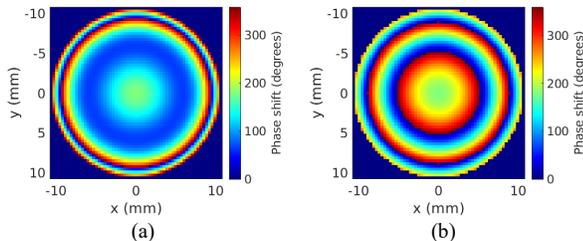

Fig. 6. The phase profiles for the (a) aperture transmitarray and (b) the focusing transmitarray.

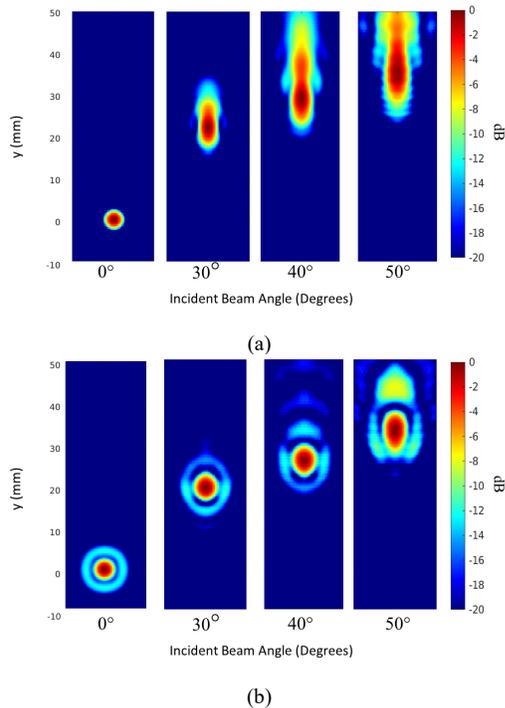

Fig. 7. Calculated intensity distributions on the image plane for different (source) incident angles. The transmitarray(s) used are (a) a conventional parabolic phase-shifting transmitarray and (b) a doublet transmitarray.

The use of a doublet lens configuration is a well-established method in classical lens optics to correct for spherical aberrations. Historically, a Schmidt plate has been used to correct spherical aberrations in optical applications, leading to the Schmidt telescope and Schmidt camera [32]. A spherical lens causes principal rays to bend less than marginal rays, resulting in misalignment that is too great for a single lens to correct [33]. A Schmidt plate corrects for that by making the marginal rays diverge while making the principal rays converge. This compensates for the unbalanced amount of bending experienced by the rays when it passes through the spherical lens. The optimized aperture transmitarray has a phase profile resembling the Schmidt plate. As such, it corrects for the aberrations introduced by the traditional focusing transmitarray, which has a phase profile akin to that of a spherical lens. The aperture transmitarray itself is also unable to operate independently, as it exists to introduce divergence and convergence that is to be 'corrected' by the focusing transmitarray. Without the focusing transmitarray, the focus spot produced by the aperture transmitarray alone would also appear defocused. While our design shares similarities with such optical elements, our transmitarrays are nevertheless more compact and lightweight than conventional optical components.

The cost function simulated across different angles for running the optimization process for 100 iterations over interplane distances, which ranged from 1 mm to 10 mm is shown in Fig. 4(b). The distance of 5 mm between the aperture and focusing transmitarray is selected after testing with various interplane distances. It is found that an interplane distance of 5 mm yields the smallest accumulated error. The optimized phase profiles for the two transmitarrays are shown in Fig. 6.

### C. Results

Fig. 6 shows the theoretical intensity distributions on the image plane calculated by the model developed in Section II for the conventional transimtarray with parabolic phase profile and the proposed doublet transmitarray architecture. Furthermore, the one-dimensional cross-section of the focal spot intensity is shown in Fig. 5 for different feed rotation angles. Fig. 7(a) clearly shows the focal spots' elongation in conventional parabolic phase-shifting transmitarrays, displaying the comatic effects. For conventional parabolic phase-shifting transmitarray, a feed displacement angle of 0°, the -3 dB spot size is 2λ. However, at larger angles such as 40° and 50°, the spot size spreads out to be more than 10λ, which is unacceptable in scanning applications. Using the doublet rectifies the situation as shown in Fig. 7(b) for doublet transmitarray, the intensity peak is far narrower and has a width of 4λ even for an incident angle as large as 50°. It demonstrates that the system has effectively corrected for the effects of focusing a source that has large oblique incident angles.

### IV. TRANSMITARRAY ELEMENT DESIGN

Fig. 8 shows the configurations of the polarization-conversion element used for the building block of the two transmitarrays. The element consists of a 45°-tilted C-shaped metallic layer sandwiched by two orthogonal wire gratings [34]. The lattice size of the unit cell is P = 0.33 mm, corresponding to 0.32λ at 290 GHz. The 0.127 mm-thick Rogers 5880 with a measured relative dielectric constant $\varepsilon_r$ = 2.3 and loss tangent tanδ = 0.004 at 290 GHz [35] is adopted as the substrate of the element. The transmitarray element can convert the incident x-polarized (x-pol) waves into the transmitted orthogonal y-polarized (y-pol) waves with high efficiency. This is enabled by the multi-reflections between the two orthogonally-polarized

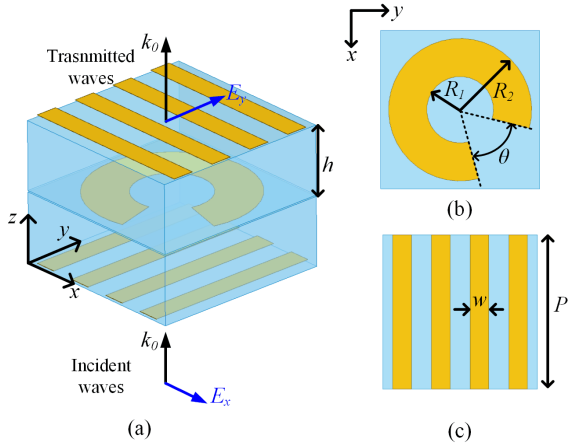

Fig. 8. Configuration of the C-shaped polarization-conversion transmitarray element. (a) 3D perspective view of the transmitarray element; (b) Top view of the middle C-shaped pattern; (c) Top view of the bottom wire grating. The dimensions are R1 = 0.07 mm, R2 = 0.15 mm, w = 0.04 mm, h = 0.127 mm.

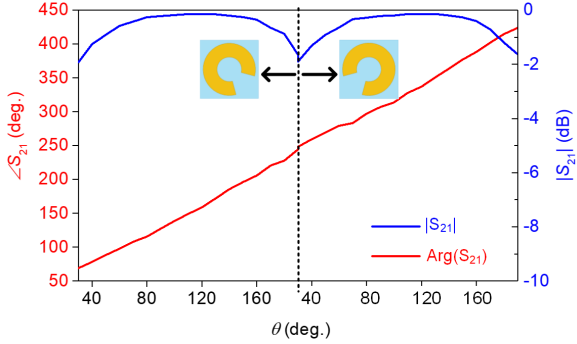

Fig. 9. Simulated transmission coefficients as a function of the split-angle of the C-shaped pattern at 280 GHz.

wire gratings and the middle *C*-shaped polarization-conversion component [36].

The split angle of the middle *C*-shaped pattern $\theta$ is tuned to change the transmission phase while other parameters are fixed. The simulated transmission coefficient as a function of the split angle at 280 GHz is shown in Fig. 9. It can be observed that the 180º transmission phase range can be observed as $\theta$ varies from 30º to 190º, with insertion loss better than -2 dB. Another 180º phase range can be achieved by simply rotating the *C*-shaped pattern along its geometric center by 90º, as shown in Fig. 8. Compared to the *C*-shaped element in [34] our unit cell can provide continuous 360º phase range by simply tuning just one geometric parameter ($\theta$) of the element.

While conventional PCB process may offer acceptable fabrication accuracy for transmitarray fabrication at H-band, micromachining technology can provide a much higher accuracy of 0.5 µm, resulting in a much smaller fabrication error. The metal used for the transmitarrays is aluminum with thickness of around 0.6 µm. The detailed micromachining process can be found in [14]. But different from using the transparent Benzocyclobutene dielectric in [14], opaque commercial Rogers 5880 is adopted as the substrate in this work, making the alignment of different metallic layers challenging during the fabrication process. To investigate the misalignment effects on the performance of the transmitarray, Fig. 10(a) shows the unit cell surrounded by periodic boundary conditions (PBC) with different kinds of offsets. Model A presents the perfectly aligned case, while B and C show the element structures after performing offsets along the *y*- and *x*-directions to the middle *C*-shaped pattern. Model D further introduces *x*-direction offset to the bottom wire grating pattern. These different models are simulated in ANSYS HFSS. The simulated transmission coefficients as a function of the split angle of the *C*-shaped pattern are shown in Fig. 9. It can be observed that models A-D share almost identical transmission curves. As a result, the introduced *C*-shaped polarization-conversion element is free of alignment problems among different metallic layers, making the microfabrication of the transmitarray straightforward. This unique and appealing property is enabled by the multi-reflection working principle of the polarization conversion element. We then modeled the whole antenna system, including the transmitarray doublet and the feed horn in ANSYS HFSS to perform the full-wave simulation. The simulated total efficiency, defined as the ratio of the radiated power to the input power of the whole antenna system, is 89%.

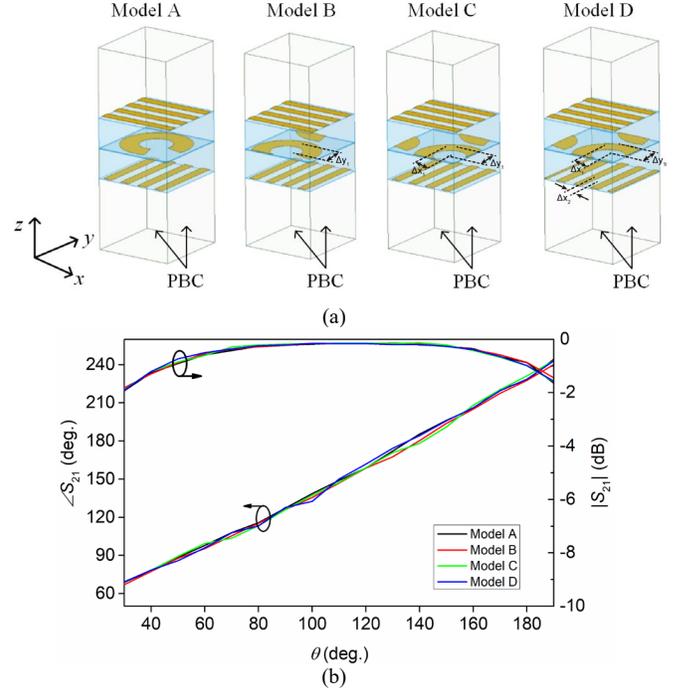

Fig. 10. (a) Configurations of the unit cell under periodic boundary conditions with different types of metallic layer offsets. The offset values are Δ$y_1$ = 0.1 mm for Model B, (Δ$x_1$, Δ$y_1$) = (0.1mm, 0.1mm) for Model C, and (Δ$x_1$, Δ$y_1$, Δ$x_1$) = (0.1mm, 0.1mm, 0.02mm) for Model C, respectively. (b). Simulated reflection coefficients as a function of split angle with different offsets.

## V. EXPERIMENTAL DEMONSTRATION

The dual-transmitarray introduced in the above sections was fabricated and measured to demonstrate the proposed approach and design. Fig. 11 shows the micro-machining transmitarray under the microscope. A picture of the experimental setup is shown in Fig. 12. A horn source is placed on a platform that traces an arc always 56 mm away from the transmitarray unit.

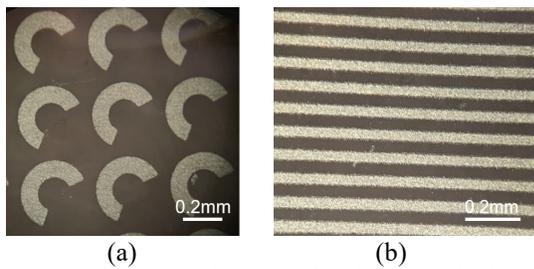

Fig. 11. Picture taken under a microscope of (a) the C-shaped polarization-conversion transmitarray element and (b) the wire grating.

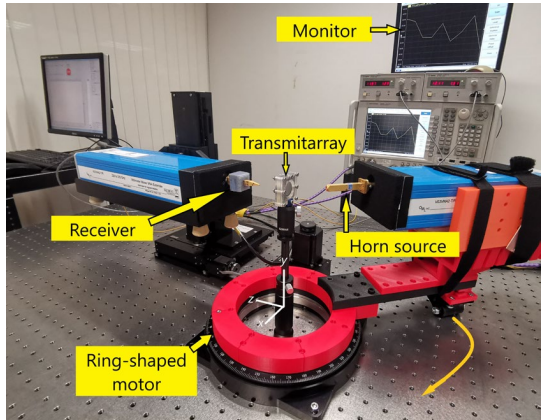

Fig. 12. Photograph of the measurement setup.

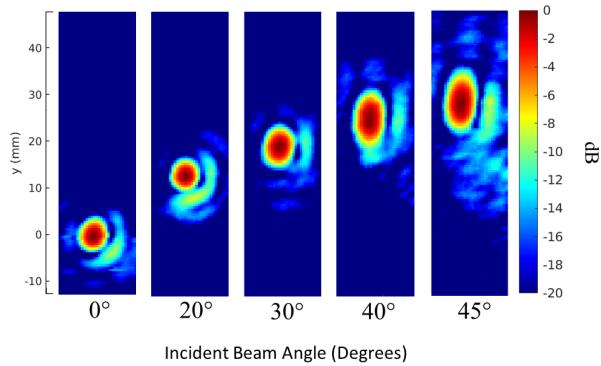

Fig. 13. Measurement results of transmitarray performance at 280 GHz.

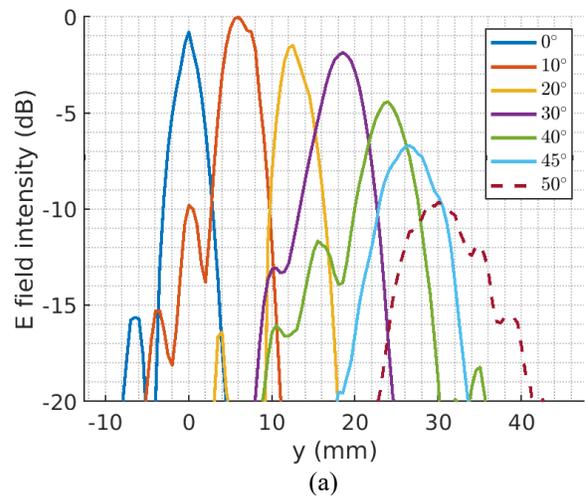

(a)

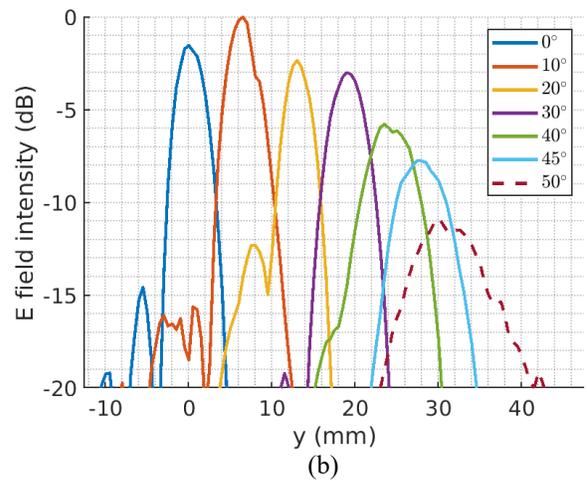

(b)

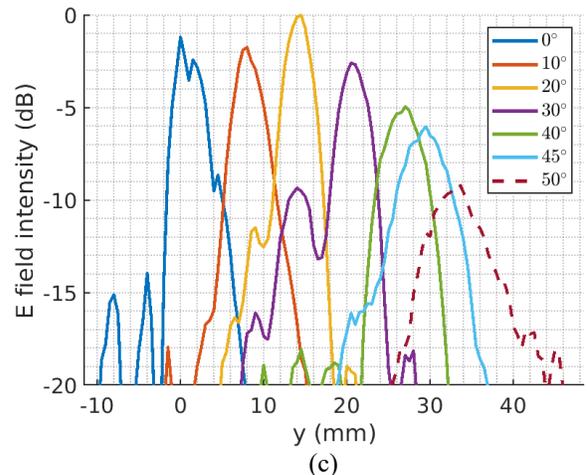

(c)

Fig.14. Measured intensity distribution in cross section at (a) 260GHz, (b) 280GHz, and (c) 300GHz.

Specifically, a circular ring-shaped high-precision servo motor is used to move the horn source. The transmitarray is placed at the center of the rotary table. The ring can be rotated, allowing the horn source to move in a circular arc at a constant radius of $z_1 = 56$mm away from the transmitarray. A receiver is placed $z_2 = 40$ mm away from the transmitarray. It is important to point out that, we utilized a bulky THz frequency extender (OML V03VNA2-T/R) as the feed source of the transmitarray for ease of measurement, as shown in Fig. 12. However, in practical applications, lightweight and small-footprint THz IC source with mm-scale size [24], [25] are available to work as the feed source of the transmitarray.

The measured power densities on the imaging plane for different incident beam angles at 280 GHz are shown in Fig. 13. It is observed that a small and scanning focusing spot can be achieved for the designed transmitarray antenna. It is noted that although only five discrete incident cases are given in Fig.13 for brevity, the transmitarray antenna can realize continuous near-field focus scanning by continuously moving the feed horn along the designed arc. Moreover, 2-D near-field focusing is also feasible by moving the feed source considering the center symmetric phase profile provided by the transmitarray doublet. The cross-sectional intensities of the focus spots for the measurement at 260 GHz, 280 GHz, and 300 GHz are displayed in Fig. 14. The measured scan loss is 6 dB up to 45° incidence at 300 GHz. The dual-transmitarray antenna can achieve wideband near-field focus scanning performance from 260 GHz to 300 GHz. The measured correction at 50° incidence is not as significant as that presented in the simulation results as shown in Fig. 5. A reason for this is that the theoretical model presented in the earlier section does not account for the diffraction effect which occurs at the edges of the transmitarray, especially for large incident angles. However, the double transmitarray can mitigate the off-axis aberration up to 45° incidence with measured near-field SLL better than -13.3 dB in $yz$-plane. Future development can be made to improve the theoretical model to consider the diffractive edge effects.

## VI. Conclusion

This paper presents a novel dual-transmitarray antenna to correct for comatic aberration introduced by traditional THz focusing systems. A propagation model was constructed and used in conjunction with an optimization algorithm to produce a design that could minimize the comatic aberration introduced by a traditional focusing method. This paper considered a setup involving a horn source that can move in a circular arc at $r = 56$ mm away from the transmitarray unit and images onto a plane located at 40 mm away. The resultant design was subsequently constructed and experimentally demonstrated. The measured results verify that the focus spots remain tight with minimum sidelobes level up to an off-axis angle of 45°. This makes the method extremely valuable in improving the focusing aspect of THz scanning systems and can be applied to create a more robust and portable system.

This work has only considered a direct system with transmitarrays, and the measurement was only performed by a 1-D scanning. As a proof of concept, it demonstrates that it is possible to design antenna systems that minimize comatic aberrations for off-axis scanning beams operating in the THz region. Such a method can easily be generalized to 2-D focus scanning. The devised system and strategy work with transmitarrays; it can be generalized with similar setups in applications involving reflectarrays, lens antennas, and reflectors. In general, such a method can be applied in systems that require tight focus spots in off-axis sources. It can also be easily modified to be used with higher or lower frequencies, such as the regime of microwave and millimeter-wave, thus making it widely applicable across various other problems. Future work can be done to apply the same strategy and methodology to improve other scanning antenna systems.

## References

[1] H. Elayan, O. Amin, B. Shihada, R. M. Shubair, and M.-S. Alouini, 'Terahertz Band: The Last Piece of RF Spectrum Puzzle for Communication Systems', *IEEE Open J. Commun. Soc.*, vol. 1, pp. 1–32, 2020, doi: 10.1109/OJCOMS.2019.2953633.

[2] M. Tonouchi, 'Cutting-edge terahertz technology', *Nat. Photonics*, vol. 1, no. 2, Art. no. 2, Feb. 2007, doi: 10.1038/nphoton.2007.3.

[3] P. H. Siegel, 'Terahertz technology in biology and medicine', *IEEE Trans. Microw. Theory Tech.*, vol. 52, no. 10, pp. 2438–2447, Oct. 2004, doi: 10.1109/TMTT.2004.835916.

[4] S. W. Smye, J. M. Chamberlain, A. J. Fitzgerald, and E. Berry, 'The interaction between Terahertz radiation and biological tissue', *Phys. Med. Biol.*, vol. 46, no. 9, pp. R101-112, Sep. 2001, doi: 10.1088/0031-9155/46/9/201.

[5] H. Quast and T. Loffler, '3D-terahertz-tomography for material inspection and security', in *2009 34th International Conference on Infrared, Millimeter, and Terahertz Waves*, Sep. 2009, pp. 1–2. doi: 10.1109/ICIMW.2009.5325639.

[6] K. B. Cooper, R. J. Dengler, N. Llombart, B. Thomas, G. Chattopadhyay, and P. H. Siegel, 'THz Imaging Radar for Standoff Personnel Screening', *IEEE Trans. Terahertz Sci. Technol.*, vol. 1, no. 1, pp. 169–182, Sep. 2011, doi: 10.1109/TTHZ.2011.2159556.

[7] P. H. Siegel, 'Terahertz technology', *IEEE Trans. Microw. Theory Tech.*, vol. 50, no. 3, pp. 910–928, Mar. 2002, doi: 10.1109/22.989974.

[8] G.-B. Wu, S.-W. Qu, and S. Yang, 'Wide-Angle Beam-Scanning Reflectarray With Mechanical Steering', *IEEE Trans. Antennas Propag.*, vol. 66, no. 1, pp. 172–181, Jan. 2018, doi: 10.1109/TAP.2017.2775282.

[9] P. Nayeri, F. Yang, and A. Z. Elsherbeni, 'Bifocal Design and Aperture Phase Optimizations of Reflectarray Antennas for Wide-Angle Beam Scanning Performance', *IEEE Trans. Antennas Propag.*, vol. 61, no. 9, pp. 4588–4597, Sep. 2013, doi: 10.1109/TAP.2013.2264795.

[10] M. Born and E. Wolf, *Principles of Optics: Electromagnetic Theory of Propagation, Interference and Diffraction of Light*. Cambridge University Press, 1999.

[11] G. R. Fowles, *Introduction to Modern Optics*. Courier Corporation, 1989.

[12] A. H. Abdelrahman, F. Yang, A. Z. Elsherbeni, and P. Nayeri, *Analysis and Design of Transmitarray Antennas*. Cham: Springer International Publishing, 2017. doi: 10.1007/978-3-031-01541-0.

[13] Payam Nayeri, Fan Yang, and Atef Z. Elsherbeni, *Reflectarray Antennas: Theory, Designs, and Applications*. Accessed: Oct. 23, 2022. [Online]. Available: https://ieeexplore.ieee.org/book/8320444

[14] S.-W. Qu, H. Yi, B. J. Chen, K. B. Ng, and C. H. Chan, 'Terahertz Reflecting and Transmitting Metasurfaces', *Proc. IEEE*, vol. 105, no. 6, pp. 1166–1184, Jun. 2017, doi: 10.1109/JPROC.2017.2688319.

[15] L. Dussopt, K. Medrar, and L. Marnat, 'Millimeter-Wave Gaussian-Beam Transmitarray Antennas for Quasi-Optical $\$\$\$ -Parameter Characterization', *IEEE Trans. Antennas Propag.*, vol. 68, no. 2, pp. 850–858, Feb. 2020, doi: 10.1109/TAP.2019.2943417.

[16] E. G. Plaza *et al.*, 'An Ultrathin 2-bit Near-Field Transmitarray Lens', *IEEE Antennas Wirel. Propag. Lett.*, vol. 16, pp. 1784–1787, 2017, doi: 10.1109/LAWP.2017.2678599.

[17] H.-T. Chou, T.-M. Hung, N.-N. Wang, H.-H. Chou, C. Tung, and P. Nepa, 'Design of a Near-Field Focused Reflectarray Antenna for 2.4 GHz RFID Reader Applications', *IEEE Trans. Antennas Propag.*, vol. 59, no. 3, pp. 1013–1018, Mar. 2011, doi: 10.1109/TAP.2010.2103030.

[18] K. Narayanasamy, G. N. A. Mohammed, K. Savarimuthu, R. Sivasamy, and M. Kanagasabai, 'A comprehensive analysis on the state-of-the-art developments in reflectarray, transmitarray, and transmit-reflectarray antennas', *Int. J. RF Microw. Comput.-Aided Eng.*, vol. 30, no. 9, p. e22272, 2020, doi: 10.1002/mmce.22272.

[19] B. Groever, W. T. Chen, and F. Capasso, 'Meta-Lens Doublet in the Visible Region', *Nano Lett.*, vol. 17, no. 8, pp. 4902–4907, Aug. 2017, doi: 10.1021/acs.nanolett.7b01888.

[20] W. Menzel, M. Al-Tikriti, and R. Leberer, 'A 76 GHz multiple-beam planar reflector antenna', in *2002 32nd European Microwave Conference*, Sep. 2002, pp. 1–4. doi: 10.1109/EUMA.2002.339453.

[21] G.-B. Wu, S.-W. Qu, S. Yang, and C. H. Chan, 'Low-Cost 1-D Beam-Steering Reflectarray With ±70° Scan Coverage', *IEEE Trans. Antennas Propag.*, vol. 68, no. 6, pp. 5009–5014, Jun. 2020, doi: 10.1109/TAP.2019.2963572.

[22] W. Menzel, D. Pilz, and M. Al-Tikriti, 'Millimeter-wave folded reflector antennas with high gain, low loss, and low profile', *IEEE Antennas Propag. Mag.*, vol. 44, no. 3, pp. 24–29, Jun. 2002, doi: 10.1109/MAP.2002.1028731.

[23] N. Llombart, R. J. Dengler, and K. B. Cooper, 'Terahertz Antenna System for a Near-Space-Rate Radar Imager [Antenna Applications]', *IEEE Antennas Propag. Mag.*, vol. 52, no. 5, pp. 251–259, Oct. 2010, doi: 10.1109/MAP.2010.5687559.

[24] H. Sherry *et al.*, 'Lens-integrated THz imaging arrays in 65nm CMOS technologies', in *2011 IEEE Radio Frequency Integrated Circuits Symposium*, Jun. 2011, pp. 1–4. doi: 10.1109/RFIC.2011.5940670.

[25] P. Hillger, J. Grzyb, R. Jain, and U. R. Pfeiffer, 'Terahertz Imaging and Sensing Applications With Silicon-Based Technologies', *IEEE Trans. Terahertz Sci. Technol.*, vol. 9, no. 1, pp. 1–19, Jan. 2019, doi: 10.1109/TTHZ.2018.2884852.

[26] J. W. Goodman, *Introduction to Fourier Optics*. Freeman & Company, W. H., 2017.

[27] D. G. Voelz, *Computational Fourier Optics: A MATLAB® Tutorial*. 1000 20th Street, Bellingham, WA 98227-0010 USA: SPIE, 2011. doi: 10.1117/3.858456.

[28] A. Grbic and R. Merlin, 'Near-Field Focusing Plates and Their Design', *IEEE Trans. Antennas Propag.*, vol. 56, no. 10, pp. 3159–3165, Oct. 2008, doi: 10.1109/TAP.2008.929436.

[29] A. Grbic, R. Merlin, E. M. Thomas, and M. F. Imani, 'Near-Field Plates: Metamaterial Surfaces/Arrays for Subwavelength Focusing and Probing', *Proc. IEEE*, vol. 99, no. 10, pp. 1806–1815, Oct. 2011, doi: 10.1109/JPROC.2011.2106191.

[30] S. R. Rengarajan, 'Scanning and Defocusing Characteristics of Microstrip Reflectarrays', *IEEE Antennas Wirel. Propag. Lett.*, vol. 9, pp. 163–166, 2010, doi: 10.1109/LAWP.2010.2045217.

[31] S. Kirkpatrick, C. D. Gelatt, and M. P. Vecchi, 'Optimization by Simulated Annealing', *Science*, vol. 220, no. 4598, pp. 671–680, May 1983, doi: 10.1126/science.220.4598.671.

[32] G. Lemaître, 'New procedure for making schmidt corrector plates', *Appl. Opt.*, vol. 11, no. 7, pp. 1630–1636, Jul. 1972, doi: 10.1364/AO.11.001630.

[33] A. Schuster, *An introduction to the theory of optics... / revised and enlarged by [Sir Arthur Schuster] and John William Nicholson.*, 3rd ed. E. Arnold, 1928.

[34] K. Medrar *et al.*, 'H-Band Substrate-Integrated Discrete-Lens Antenna for High Data Rate Communication Systems', *IEEE Trans. Antennas Propag.*, vol. 69, no. 7, pp. 3677–3688, Jul. 2021, doi: 10.1109/TAP.2020.3044382.

[35] X. Ruan and C. H. Chan, 'Terahertz free-space dielectric property measurements using time- and frequency-domain setups', *Int. J. RF Microw. Comput.-Aided Eng.*, vol. 29, no. 9, p. e21839, 2019, doi: 10.1002/mmce.21839.

[36] G.-B. Wu, K. F. Chan, K. M. Shum, and C. H. Chan, 'Millimeter-Wave and Terahertz OAM Discrete-Lens Antennas for 5G and Beyond', *IEEE Commun. Mag.*, vol. 60, no. 1, pp. 34–39, Jan. 2022, doi: 10.1109/MCOM.001.2100523.